\documentclass[aps,prl,twocolumn]{revtex4}
\usepackage{latexsym}
\usepackage{amsmath}
\usepackage{amssymb}
\usepackage{eufrak}
\usepackage{euscript}
\usepackage{epsfig,graphics,graphicx}% Include figure files
\usepackage{bm}% bold math 
\usepackage{color}
\usepackage{slashed}

\def\be{\begin{eqnarray}}
\def\ee{\end{eqnarray}}

\begin{document}

\title{Quark mass and isospin dependence of the deconfining critical temperature}

\author{E. S. Fraga$^1$, L. F. Palhares$^1$, and C. Villavicencio$^2$}

\affiliation{$^1$Instituto de F\'\i sica, Universidade Federal do Rio de Janeiro, 
C.P. 68528, Rio de Janeiro, RJ 21941-972, Brazil \\
$^2$Departamento de F\'\i sica, Comisi\'on Nacional de Energ\'\i a At\'omica, 
(1429) Buenos Aires, Argentina
}

\begin{abstract}
We describe the deconfining critical temperature dependence on the pion mass and on 
the isospin chemical potential in remarkably good agreement with lattice data.
Our framework incorporates explicit dependence on quark 
masses, isospin and baryonic chemical potentials for the case of two flavors
through ingredients from well-known high- and low-energy theories.
In the low-energy sector, the system is described by a minimal chiral perturbation 
theory effective action, corresponding to a hot gas of pion quasiparticles and heavy 
nucleons. For the high-temperature sector we adopt a simple extension of the fuzzy 
bag model. We also briefly discuss the effects of mass asymmetry and 
baryon chemical potential.
\end{abstract}

%\date{\today}

\maketitle

%%%%%%%%%%%%%%%%%%%%%%%%%%%%%%%%%%%%%%%%%%

{\it Introduction}. The phase diagram of quark matter has been the object of intense investigation 
during the last years, and yet several open questions within the thermodynamics of 
strong interactions still remain unsolved \cite{Kogut:2004su,Yagi:2005yb}. In this quest, 
Lattice QCD represents the main non-perturbative approach within the full 
theory \cite{Laermann:2003cv}, always complemented by effective models \cite{Stephanov:2007fk}. 

In this article we investigate the effects of finite quark masses
and isospin
number on the equation of state of hot and dense strongly interacting matter 
and on the deconfining phase transition within a framework inspired by chiral 
perturbation theory ($\chi$PT) and lattice results for the pressure and the trace anomaly.
The
setting we propose is simple and {\it completely} fixed by vacuum QCD properties
(measured or simulated on the lattice) and lattice simulations of finite temperature QCD.
More explicitly, there is no fitting of mass or isospin chemical potential dependence at all.
Nevertheless,
our findings for the behavior of the critical temperature as a function of
{\it both}
the pion mass and 
the isospin chemical potential are in remarkably good agreement with lattice data.
It is crucial to note that several detailed studies of chiral models failed to 
describe $T_c(m_{\pi})$ \cite{Berges:1997eu,Dumitru:2003cf,Braun:2005fj}, while 
Polyakov-loop models, whose predictions can be fitted to the lattice points for
$T_c(m_{\pi})$ \cite{Dumitru:2003cf}, cannot at the present form address isospin effects.
In this approach, we can also investigate
in a straightforward manner
the effects of quark-mass asymmetry and nonzero baryon chemical
%
% potential.
potential, physical cases in which the Sign Problem develops, constraining
systematic lattice studies. The predictions within our framework for these regimes
which are not yet fully probed by lattice QCD are left for a longer publication
(for preliminary results, see \cite{Palhares:2008px}).
Throughout the present paper, the baryon chemical potential is fixed to zero.
It is not our aim in this short paper to provide a complete description of
the full richness of the QCD phase diagram. Still, we present a description
of the dependence of the critical deconfining temperature on the quark-average mass
and on the isospin chemical potential {\it simultaneously}. 
% To the best of our
% knowledge, this is the first successful approach when compared to lattice data.

Small quark masses and a nonzero baryon chemical potential have 
always represented a major challenge for lattice simulations. Presently, 
although viable lattice sizes still prevent extensive and precise 
studies with realistic quark masses, and the Sign Problem considerably restricts 
the applicability of Monte Carlo simulations to the description of chemically 
asymmetric media, Lattice QCD is starting to provide results with smaller 
quark masses, and probing a larger domain at finite chemical 
potential \cite{Fodor:2007sy,Cheng:2007jq,Philipsen:2008gf}. For the latter, it is 
believed that much can be learned from simulations of realizations of QCD 
that avoid the Sign Problem, such as those with vanishing baryon chemical potential 
and finite isospin density, which has a positive fermionic determinant.

QCD at finite isospin density is certainly, but not only, a playground 
to test numerical approaches to the case of finite baryon density on the 
Lattice \cite{Kogut:2004zg,deForcrand:2007uz}. It is also part of the 
physical phase diagram for strong interactions, and exhibits a very rich 
phenomenology \cite{Son:2000xc}. It has been under careful study 
during the last years, both theoretically and experimentally, with a clear identification 
of certain phenomena that depend directly on the isospin asymmetry 
of nuclear matter at intermediate-energy heavy ion experiments (see Ref.  \cite{Li:2008gp} 
and references therein).

Nevertheless, theoretical and phenomenological studies often focus on the 
chiral limit of QCD, putting aside effects from finite quark masses, and 
on isospin symmetric
%
% symmetric, baryon-free 
% 
hot matter, mainly stimulated by the 
physical scenario found in current high-energy heavy ion collision 
experiments \cite{QM} and the quark-gluon plasma \cite{Rischke:2003mt}.
Some exceptions are, however, chiral model analyses of pion-mass dependence
of the finite temperature transition (e.g., \cite{Berges:1997eu,Dumitru:2003cf,Braun:2005fj}),
which fail completely to reproduce the lattice behavior, and Polyakov-loop
models predictions as mentioned above. At finite isospin chemical potential,
though there are phase-diagram investigations in PNJL models \cite{Mukherjee:2006hq},
none of them has addressed the critical temperature dependence on the isospin chemical
potential in particular.
% 
% 
% {\bf\red Discutir o que j\'a existe... exceptions...}

To investigate the effects of nonzero quark 
% 
% masses, baryon chemical potential, 
masses
and isospin asymmetry on the deconfining transition, we build an effective theory that 
combines ingredients from $\chi$PT in the low-energy sector with the 
phenomenological fuzzy bag model at high energy. The high-energy regime is described 
perturbatively by two-flavor QCD with massive quarks and explicit isospin symmetry breaking. 
Nonperturbative (confinement) effects at this scale are incorporated through a fuzzy bag 
description \cite{Pisarski:2006hz} with coefficients extracted from lattice data. For the 
low-energy sector, we adopt an effective action inspired by $\chi$PT that 
exhibits the same structure of symmetries contained in the high-energy theory \cite{Loewe:2002tw}.
The quasi-nucleon degrees of freedom described in this regime seem to be crucial for
understanding the pion-mass dependence of the deconfining temperature.
The definition of parameters, as well as masses, is such that variations in the deconfined sector 
are totally consistent with variations in the confined one, which guarantees the bookkeeping in 
the different degrees of freedom % and quasiparticles 
that are present in the description.

By matching the two branches of the equations of state, corresponding to the
high and low temperature regimes, we of course obtain a first-order transition.
Recently improved Lattice QCD calculations, with almost realistic quark masses,
seem to indicate a crossover instead \cite{Cheng:2007jq}. On the other hand, from the experimental
standpoint a weakly first-order transition is not ruled out, and in fact corresponds
to the scenario adopted in very successful hydrodynamic calculations \cite{QM}.
Although this is a crucial question for the understanding of the phase structure
of QCD, it is essentially of no consequence to the analysis we undertake.
The value of the critical temperature we obtain is $\sim 5-10 \%$ different from
current values extracted from the Lattice \cite{Cheng:2007jq}\footnote{Currently, 
there is still no agreement between different groups
regarding the value of the critical temperature (cf. Ref. \cite{Fodor:2007sy}). This
important issue will hopefully be settled in the near future.
By convenience, we choose to compare with the set of results
that was used in previous tests of other effective models to which
we also compare our results.}, which is always the case in effective field theory approaches
and of no harm to our analysis, either. Our concern is with providing a good description of
the behavior of the critical temperature for increasing values of quark masses and the
isospin chemical potential.

%%%%%%%%%%%%%%%%%
{\it Low-energy sector}. The physical setting for the low-energy regime of strong interactions is 
that of a system of heavy nucleons in the presence of a hot gas of pions whose masses are 
already dressed by corrections from temperature, isospin chemical potential and 
quark masses. The effective Lagrangian reads 
$\mathcal{L}_{eff} = \mathcal{L}_{N}+\mathcal{L}_{\pi}+\mathcal{L}_{\pi N}$, where
\be
\mathcal{L}_{N} &=& \overline{N} \left(
i\slashed{\partial}-M_N+\frac{1}{2}\mu_I\gamma_0 \tau^3+\frac{3}{2} \mu_B \gamma_0
\right) N \, ,
\label{LN}
\ee
\be
\mathcal{L}_{\pi} &=&
(\partial_{\mu}-ih\delta_{\mu 0})\pi^+(\partial_{\mu}+ih\delta_{\mu 0})\pi^-
-\overline{m}_{\pm}^2\pi^+\pi^- \nonumber \\
&+&\frac{1}{2}\left[ (\partial \pi^0)^2 -\overline{m}_0^2(\pi^0)^2 \right]
\, , \label{Lpi}
\ee
\be
\mathcal{L}_{\pi N} &=&\frac{g_A}{f_{\pi}}\overline{N} i \gamma_5 \left(
\slashed{\partial}\pi-\frac{1}{2}\mu_I\gamma_0 [\tau^3,\pi] 
\right) N \, ,
\ee
where $g_{A}$ is the axial vector current coefficient of the nucleon, which accounts for 
renormalization in the weak decay rate of the neutron, $f_{\pi}$ is the pion decay constant, 
and $h$ is a function of temperature and isospin chemical potential. Nucleons are represented 
by $N= (p,n)$ with $p,n$ being the proton and neutron spinors, respectively, and have a mass matrix 
$M_N=\textrm{diag}(M_p,M_n)=\textrm{diag}(M-\delta M,M+\delta M)$. This corresponds to the 
${\cal O}(P)$ nucleon chiral Lagrangian \cite{Gasser:1987rb}, but considering mass corrections at 
zero temperature and chemical potential, and the coupling to dressed pions.

The effective (dressed) masses of the pions $\pi^0=\pi^3$ and 
$\pi^{\pm}=\frac{1}{\sqrt{2}}(\pi^1\mp i\pi^2)$, which depend on temperature $T$, isospin chemical 
potential $\mu_{I}$ and mass asymmetry $\delta m=(m_d-m_u)/2$, are denoted, respectively, by 
$\overline{m}_0=\overline{m}_0(T,\mu_I,\delta m)$ and 
$\overline{m}_{\pm}=\overline{m}_{\pm}(T,\mu_I)$ 
\footnote{$\mu_{I}$ and $\mu_{B}$ can be written in terms of the chemical 
potentials for the up and down quarks. Here we adopt the following convenient 
definition: $\mu_I=\mu_u-\mu_d$ and $\mu_B=\mu_u+\mu_d$. 
($\mu_{B}$ is also frequently defined with an overall factor $2/3$.) 
We also use the customary notation $\pi\equiv \pi^a\tau^a$, with $\tau^a$ being the Pauli matrices.}
and their explicit expressions were calculated in Refs. \cite{Loewe:2002tw}. 
In the so-called first phase, a regime in which $|\mu_{I}|<m_{\pi}$, 
they have the form 
\be
m_{\pi^0}&=& \overline{m}_0 
= m_{\pi}\left[ 1+\frac{1}{2}\alpha\sigma_{00} \right] \, ,
\label{m0}
\\
m_{\pi^{\pm}} &=&
\overline{m}_{\pm}\mp h
= m_{\pi}\left[ 1+\frac{1}{2}\alpha\sigma_1 \pm\frac{1}{2}\alpha\frac{\sigma_0}{m_{\pi}} \right]\mp \mu_I \, ,
\label{mpm}
\ee
up to first order in $\alpha=(m_{\pi}/4\pi f_{\pi})^{2}$. Here, $\sigma_{00}$, $\sigma_0$, $\sigma_1$, 
and $h$ are functions of temperature, isospin chemical potential and quark masses \cite{Loewe:2002tw}. 

In the second phase ($|\mu_{I}|>m_{\pi}$), a condensation of pions occurs, and a superfluid 
phase sets in \footnote{Here we consider the $\pi^-$ condensation taking $\mu_I=-|\mu_I|$.}. 
In this new phase, in order to reestablish the vacuum structure, a chiral rotation is produced  due to 
the isospin symmetry breaking. All this produces a pion mixing, and the nucleons also couple 
in a different way. The degrees of freedom do not correspond anymore to pions, but we can still call 
them quasi-pions since their masses in the two phases match at the transition point. 
%match with the masses in the normal phase. 
The tree level masses do not have the shape as in the equations above. Instead, $m_{\pi^0}=|\mu_I|$, 
$m_{\pi^-}=0$, and $m_{\pi^+}=\mu_I\sqrt{1+3(m_\pi/\mu_I)^4}$ \cite{Son:2000xc}. However, it is 
possible to treat this phase with a simple approximation near the superfluid phase transition.
In the regime in which $|\mu_{I}|\gtrsim m_{\pi}$, the natural expansion parameter is given by 
$s^{2}=1-m_{\pi}^{4}/\mu_{I}^{4}$ \cite{Loewe:2002tw}, after scaling all the parameters by $|\mu_I|$. 
The result of the first terms in this expansion ($s^2=0$) provides the same equations as in the normal 
phase, just replacing $m_\pi$ by $|\mu_I|$. Strictly speaking, this is valid only for values of $|\mu_I|$ 
very close to $m_\pi$, i.e. $|\mu_I|\lesssim \sqrt{8/7} ~m_{\pi}$ (for a more detailed discussion cf.
\cite{Loewe:2002tw}). For simplicity, we also apply this results for slightly higher values of the isospin 
chemical potential
\footnote{
For higher values of $|\mu_I|$ one needs to consider more terms, not only in the expansion of the 
pion Lagrangian but also in the coupling with the nucleons, due to the appearance of the pion 
condensate. These terms will also be present at the perturbative QCD level via 
the chiral rotation. The case in which $|\mu_{I}| \gg m_{\pi}$ can be explored, nevertheless, 
using $m_{\pi}^{2}/\mu_{I}^{2}$ as an expansion parameter, even though 
the validity of the whole treatment is restricted to $|\mu_{I}|$ smaller than the 
$\eta$ or $\rho$ masses \cite{Son:2000xc}. 
}.
These results in $\chi$PT are confirmed by a NJL analysis \cite{He:2005nk}.

The direct effect of the baryonic chemical potential in the pure pion quasiparticle 
gas is omitted since, without considering gluonic corrections, it appears as an anomalous 
term in the ${\cal O}(P^4)$ chiral Lagrangian \cite{AlvarezEstrada:1995mh}, and will be present 
only in two-loop corrections according to power counting. For very large values of 
$\mu_{B}$, one has in principle to incorporate effects from the color superconductivity gap in the calculation 
of meson masses in an effective theory near the Fermi surface \cite{Alford:2007xm,Beane:2000ms}. 
In the present analysis, we treat the case $\mu_B=0$.

The nucleon masses, $M_p=M-\delta M$ and $M_n=M+\delta M$, are dressed by leading-order 
contributions in zero-temperature baryon $\chi$PT. Using the results from 
Ref. \cite{Procura:2006bj}, for the isospin symmetric case with explicit chiral symmetry breaking, 
and Ref. \cite{Beane:2006fk}, which includes explicit isospin breaking effects, we have 
(neglecting terms $\sim m_q^2$, $\sim m_q^2\log(m_q)$, and of higher order in $m_q$):
\be
M(m)&=& M_0 +2~\gamma_1~m+2^{3/2}~\gamma_{3/2}~m^{3/2} \, ,
\\
\delta M(\delta m)&=& 2~\gamma_1^{\textrm{asymm}}~\delta m
\, ,
\ee
$M_0$ being the nucleon mass in the chiral limit, $m=(m_u+m_d)/2$ the average quark mass, 
and \footnote{To relate the pion mass (in the isospin symmetric case) with the quark masses,
we use the Gell-Mann -- Oakes -- Renner relation: 
$m_{\pi}^2f_{\pi}^2=-(m_u+m_d)\langle\bar q q\rangle=2m(-\langle\bar q q\rangle)$.}
\be
\gamma_1 &=& \frac{-4~c_1}{f_{\pi}^2}~(-\langle\bar q q\rangle) \,,
\\
\gamma_{3/2} &=&-~\frac{3g_A^2}{32\pi f_{\pi}^5}~(-\langle\bar q q\rangle)^{3/2} \, ,
\\
\gamma_1^{\textrm{asymm}} &=&
\frac{2\bar\alpha-\bar\beta}{3}~\frac{(-\langle\bar q q\rangle)}{f_{\pi}^2}
\, .
\ee
Here, all parameters and coefficients are fixed to reproduce properties of the
QCD vacuum either measured or extracted from recent lattice simulations.
Explicitly, $\langle\bar q q\rangle = -(225~\textrm{MeV})^3$ is the (1-flavor) chiral condensate 
in the chiral limit \cite{Chiu:2003iw} and, from Table I in Ref. \cite{Procura:2006bj},
$M_0 = (0.882 \pm 0.003)~\textrm{GeV}$, 
$c_1 = (-0.93\pm0.04)~\textrm{GeV}^{-1}$, 
$g_A = 1.267$, and 
$f_{\pi} = 92.4~\textrm{MeV}$, so that  
$\gamma_1 = 4.9630 \pm 0.2135$, and 
$\gamma_{3/2} = -0.273424~\textrm{MeV}^{-1/2}$.
Finally, from Table 3 in Ref. \cite{Beane:2006fk}, one can extract 
$(2\bar\alpha-\bar\beta)/3$. Converting from lattice units to GeV 
($1~(\textrm{lattice units})=b=0.125~$fm), in the case of the $O(m_q)$ fit, we arrive at 
$\gamma_1^{\textrm{asymm}}=0.16734 \pm 0.07858$. 
This fixes the dispersion relation satisfied by the proton and neutron as 
\be
E_{\textrm{p/n}} ({\bf p}) &=& 
\sqrt{{\bf p}^2+(M \mp \delta M)^2} +\frac{3}{2}\mu_B\pm \frac{1}{2}\mu_I \, ,
\ee
where ${\bf p}$ is the 3-momentum and the antiparticle dispersion relations are obtained
from the ones above by the substitution $\mu_i \mapsto -\mu_i$.

%%%%%%%%%%%%%%%%%
{\it High-energy sector}. 
The fuzzy bag model has been proposed by Pisarski \cite{Pisarski:2006hz} as a phenomenological
parameterization of the equation of state to account for the plateau in the trace anomaly 
normalized by $T^2$, $(\epsilon-3p)/T^2$, observed in lattice results above the critical temperature.
Besides the usual MIT-type bag constant, the total pressure for QCD in this model has also a 
non-perturbative contribution $\sim T^2$ \footnote{The underlying theoretical framework is that 
of an effective theory of Wilson lines and their electric field.}: 
$p_{\textrm{deconf}}(T) \simeq p_{\textrm{pQCD}}(T)-B_{\textrm{fuzzy}}~T^2-B_{\textrm{MIT}}$.
The trace anomaly associated with this equation of state, assuming that $p_{\textrm{pQCD}}\sim T^4$, 
is then $\epsilon-3p  =  2 B_{\textrm{fuzzy}}~T^2 + 4B_{\textrm{MIT}}$.

Recently, a similar parameterization has been used to fit lattice results for
the trace anomaly at high temperatures \cite{Cheng:2007jq}, $T>1.5~T_c \approx 300~$MeV , yielding 
%\footnote{In the regime $T_c<T<1.5~T_c$, the authors of Ref. \cite{Cheng:2007jq} identify a 
%strongly non-perturbative regime, which is not included in the fuzzy-type fit.}
%
\be
\left( \frac{\epsilon-3p}{T^4} \right)_{\textrm{high T}} &=&
\frac{3}{4}~b_0~g^4+\frac{b}{T^2}+\frac{c}{T^4}
\, , \label{latticefuzzyfit}
\ee
with coefficients $b$ and $c$ given in Table VIII of Ref. \cite{Cheng:2007jq}. 
Notice that the first term in Eq. (\ref{latticefuzzyfit}) comes from a $O(\alpha_s^2)$ perturbative
contribution to the pressure and is important for the fit only at very high temperatures.
Hence, we neglect this term in what follows, obtaining the following values for the bag coefficients: 
$B_{\textrm{fuzzy}}= 0.05~$Ge$V^2$ and 
$B_{\textrm{MIT}}=0.006~$GeV$^4$. 

In our effective theory, we adopt, phenomenologically, a simple extension of the fuzzy bag 
equation of state which includes the influence of finite chemical potentials and masses in 
the perturbative pressure, neglecting for simplicity non-perturbative contributions due to the 
finite quark chemical potentials $\mu_f$, so that: 
$p_{\textrm{deconf}} \simeq p_{\textrm{pQCD}}(T,{\mu_f},m_{f})-
B_{\textrm{fuzzy}}~T^2-B_{\textrm{MIT}}$.

%%%%%%%%%%%%%%%%%%%%%%%%%%%%%%%%%%%%%%%%%%
{\it Results for the critical temperature}
As detailed above, we constructed a model using results of well-studied theories and choosing carefully
the relevant ingredients to study mass and isospin number effects. Both low- and high-energy sectors 
are completely fixed and now we turn to the determination of the prediction of this model for the
behavior of the deconfining critical temperature.
From our results for the massive free gas contribution of the pQCD pressure in the fuzzy bag model 
at finite temperature, isospin and baryon number, and the free gas pressure of quasi-pions and 
nucleons in the low-energy regime, the critical temperature and chemical potential for the deconfining 
phase transition are extracted by maximizing  the total pressure. The validity of our approach is, of course,
restricted by the scale of $\chi$PT: e.g. for $m_{\pi}\approx m_{\rho}\approx 770$ MeV, the expansion
parameter in $\chi$PT is $\alpha\approx 0.45$, so that the extension of the predictions to $m_{\pi}\lesssim 1$ GeV
is justified \footnote{A rough estimate of the error in the effective masses of the low-energy sector
of our approximation for $m_{\pi}\sim 1$ GeV yields $\sim 30\%$.}.

\begin{figure}[htb]
\vspace{0.43cm}
\includegraphics[width=7.7cm]{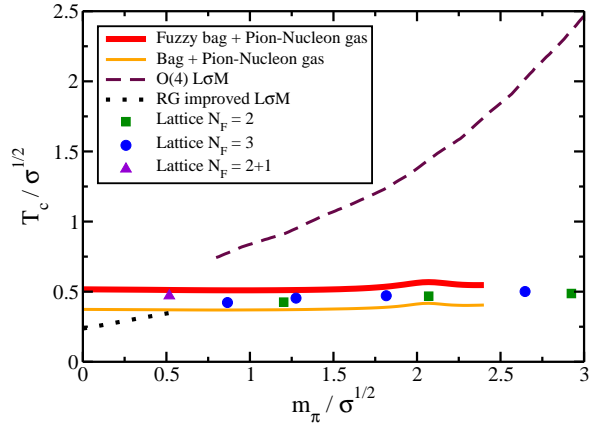}
\caption{Critical temperature as a function of $m_\pi$ normalized by 
the string tension $\sqrt{\sigma}=425~$MeV. Our results are compared to 
lattice data \cite{Karsch:2000kv,Cheng:2007jq} and other approaches \cite{Dumitru:2003cf,Berges:1997eu}.}
\label{Tc-mpi}
\end{figure}
%
%\hspace{2cm}

The pion mass dependence, or equivalently the quark average mass dependence, of the 
critical temperature is displayed in Fig. \ref{Tc-mpi} for vanishing chemical potentials 
with the temperature and the pion mass normalized by the square root of the string tension. 
Our curves stop at a point from which our $\chi$PT approach clearly 
breaks down. The results of our framework are compared to lattice data from Refs. \cite{Karsch:2000kv} 
($N_F=2,3$) and \cite{Cheng:2007jq} ($N_F=2+1$)
and two other phenomenological treatments: the $O(4)$ linear sigma model \cite{Dumitru:2003cf} 
and a renormalization group improved computation \cite{Berges:1997eu} 
(cf. also Ref. \cite{Braun:2005fj} that discusses the quark-meson model using the proper-time 
renormalization group approach). The approximate mass 
independence observed in the lattice data is very well reproduced within our framework, while 
the other descriptions tend to generate a qualitatively different behavior. 
This feature is yet another indication that the functional dependence of $T_c(m_{\pi})$
requires confinement ingredients to be reproduced, being incompatible with a phase transition dictated
by pure chiral dynamics. This argument goes in the same direction of Ref. \cite{Dumitru:2003cf},
the main difference here being the fact that we construct the mass dependence from the
$m_{quark}=0$ limit, with the heavy quasi-nucleons as the key new element at low-energies.
Moreover, our results are not strongly sensitive to the choice between the fuzzy bag model and the 
usual MIT bag model. The critical values for the MIT bag model are systematically lowered, but 
the qualitative behavior is not altered, as illustrated in Fig. \ref{Tc-mpi}. This indicates 
that a consistent treatment of the quark mass dependence connecting both perturbative regimes of 
energy is probably the essential ingredient to describe this observable.

% \vspace{.3cm}

%
\begin{figure}[htb]
\vspace{0.6cm}
\includegraphics[width=7.7cm
%, angle=270
]{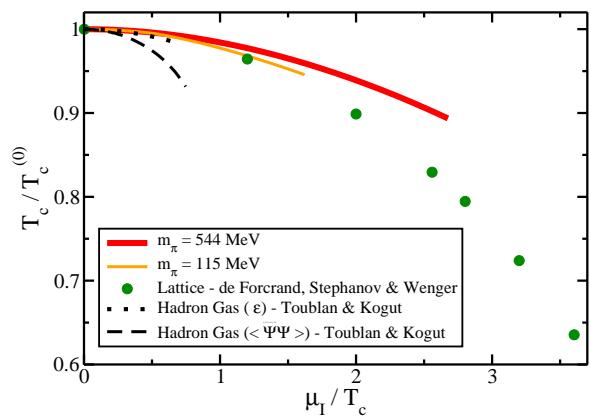}
\caption{Critical temperature as a function of $\mu_I$ for two different 
values of the pion mass. Our results are compared to lattice data \cite{deForcrand:2007uz}
and other approaches \cite{Toublan:2004ks}.}
\label{Tc-muI}
\end{figure}

In Fig. \ref{Tc-muI}, the critical temperature is plotted as a function of the isospin 
chemical potential. The critical temperature is normalized by its value in the absence 
of chemical asymmetry, whereas the isospin chemical potential is normalized by the critical 
temperature itself. The curves obtained within our framework are represented by solid lines. 
Once again our results are in good agreement with lattice computations \cite{deForcrand:2007uz}, 
even though the curve that is closer to the lattice points corresponds to a small vacuum pion mass, 
which is not the situation simulated on the lattice \cite{deForcrand:2007uz}. Our curves for different values of 
$m_{\pi}$ start to disagree more appreciably for $\mu_{I}> 2 T_{c}$. Recall that our treatment 
is valid up to isospin chemical potentials that are larger but not much larger than the 
pion mass, and contains the effects from pion condensation for $\mu_{I}> m_{\pi}$. 
Fig. \ref{Tc-muI} also exhibits, for comparison, results using the hadron resonance 
gas model \cite{Toublan:2004ks} that appear to depart from the lattice data at a much lower 
value of the isospin chemical potential. The two curves are produced by two different methods to 
determine the critical temperature: the dotted curve is obtained from the observation that the 
deconfined phase emerges at a constant energy density, whereas the dashed one uses the fact 
that the quark-antiquark condensate for the light quarks almost disappears at the quark-hadron
transition \cite{Toublan:2004ks}. Similarly to what we observe for the quark-mass dependence
of $T_c$, it is clear from Fig. \ref{Tc-muI} that plain chiral considerations render the 
largest discrepancies for the behavior of $T_c(\mu_I)$ as compared to lattice data.

A up-down quark mass imbalance, characterized by  the relative difference in quark mass, 
which is in principle not much smaller than one, tends to increase 
the critical temperature, though by a quantitatively small amount, as expected. The value of the 
critical baryonic chemical potential, beyond which matter is deconfined, also seems to increase 
with the vacuum pion mass. Detailed results in these and other thermodynamic observables will 
be presented in a longer publication (see also \cite{Palhares:2008px}).

%%%%%%%%%%%%%%%%%%%%%%%%%%%%%%%%%%%%%%%%%%
{\it Final remarks}. 
We constructed a frugal effective framework selecting features from established theories
that are relevant in the determination of the quark-mass and chemical-asymmetry dependence of the deconfining transition.
% And yet from 
From a simple free gas calculation of the equation of state, we found surprisingly good agreement 
with different lattice data for the behavior of the critical temperature for the deconfining transition 
with masses and the isospin chemical potential, indicating that the model captures the essential 
features brought about by the inclusion of those effects. 
It should be emphasized that the framework provides simultaneous predictions for these two functional dependences
of the critical temperature, $T_c(m_\pi)$ and $T_c(\mu_I)$, with only one set of coefficients
completely determined by observed QCD vacuum properties and lattice simulations at finite temperature and 
zero chemical potentials and fixed quark masses. As far as we are aware of, this is the first framework
to well-reproduce lattice data.
Our description of $T_c(m_\pi)$ reinforces, through a completely 
different approach, the discussion in Ref. \cite{Dumitru:2003cf} in which the 
quark-mass dependence of the finite temperature transition on the lattice is shown 
to be compatible with results from slightly perturbed Polyakov-loop models, contrasting 
with the failure of different chiral models. Confinement properties seem to influence strongly
the mass-dependence of the QCD transitions observed on the lattice.
Moreover, the approximate framework built in the present paper, constituted of selected ingredients from
well-kown high- and low-energy theories, is able to access physical settings including quark-mass 
asymmetry and finite baryon chemical potential, regimes in which lattice simulations encounter 
severe difficulties due to non-positive definite fermion determinants.
Therefore, this model provides a pragmatic tool to investigate the role played by nonzero 
quark masses and chemical potentials, also going beyond the free gas approximation, complementing
in a healthy direction results from other model approaches and lattice simulations. Since 
the (mass-symmetric) isospin chemical potential regime does not suffer from the Sign problem, the road is open 
for detailed Monte Carlo studies and further comparisons. On the experimental side, 
intermediate-energy experiments in nuclear physics are providing data and observables 
that are sensitive to chemical asymmetry, whereas high-energy heavy ion collisions to 
start soon at RHIC-BES and FAIR-GSI will probe a region of the phase diagram of QCD where 
effects from $\mu_{B}$ become important. Here we have focused on the influence of finite 
quark masses and isospin chemical potential. Predictions of this framework for the 
finite $\mu_B$ regime will be reported in the near future. %
This work was partially supported by ANPCyT, CAPES, CNPq, FAPERJ, and FUJB/UFRJ.
%

%%%%%%%%%%%%%%%%%%%%%%%%%%%%%%%%%%%%%%%%%%
%%%%%%%%%%%%%%%%%%%%%%%%%%%%%%%%%%%%%%%%%%

\end{document}